\documentstyle[12pt]{article}
\def\v{\vskip 2mm}

\def\r{\hangindent=1pc  \noindent}
\def\ref{\r}
\def\kms{km s$^{-1}$}
\def\deg{$^\circ$}
\def\Vlsr{V_{\rm lsr}}
\def\Msun{M_{\odot \hskip-5.2pt \bullet}}

\def\Deg{^\circ}
\def\lv{$(l, V)$}
\def\kluwer{Kluwer Academic Publishers, Dordrecht}
\begin{document}
\pagenumbering{arabic}
\title{ Radio Continuum and Molecular Gas in the Galactic Center}

\author{Yoshiaki\,Sofue\\
Institute of Astronomy, University of Tokyo,
Mitaka, Tokyo 181, Japan \\E-mail: sofue@mtk.ioa.s.u-tokyo.ac.jp}

\maketitle

\begin{abstract}
Nonthermal radio emission
in the galactic center reveals a number of vertical structures 
across the galactic plane, which are
attributed to poloidal magnetic field and/or energetic outflow.
Thermal radio emission comprises star forming regions distributed
in a thin, dense thermal gas disk.
The thermal region is associated with dense molecular gas disk, in which
the majority of gas is concentrated in a rotating molecular ring.
Outflow structures like the radio lobe is associated
with rotating molecular gas at high speed, consistent with a twisted
magnetic cylinder driven by accretion of a rotating gas disk.
\footnote{To appear in proceedings of Nobel Symposium 98:
``Barred galaxies and circumnuclear activity'',
Saltsjvbaden, Stockholm Obs., 30 Nov - 3 Dec 1995,
ed. Aa.Sandqvist }

\end{abstract}

\section{Radio Continuum Emission}

\subsection{Flat Radio Spectra}

The radio emission from the Galactic Center is a mixture of thermal
and nonthermal emissions.
The conventional method to investigate the emission mechanism is to study 
the spectral index, either flat (thermal) or steep (nonthermal).
Howerver, the spectral index in the central $3\Deg$ region has been
found to be almost everywhere flat (Sofue 1985), 
even in regions where strong linear polarization  has been detected.
Therefore, a flat spectrum observed near the galactic center can no longer
be taken as an indicator of thermal emission. 

\subsection{Infrared-to-Radio Intensity Ratio}

Separation of thermal and nonthermal radio emission can be done by comparing 
far-IR (e.g. 60 $\mu$m) and radio intensities (both in Jy/str): 
thermal (HII) regions  have high IR-to-radio ratio, 
$R=I_{\rm FIR}/I_{\rm R}\simeq 10^3$, while 
 nonthermal regions  have small IR-to-radio ratio, $R < 0 \sim 300$.
Using this method, thermal and nonthermal emission 
regions have been distinguished in a wide area (Reich et al 1987).
The region near the galactic plane is dominated by thermal emission and many 
HII regions like Sgr B2.
These regions are closely associated with dense molecular clouds 
 related to star formation in the clouds.
On the other hand we find that many of the prominent features like the Radio 
Arc, Sgr A and regions high above the galactic plane including the 
Galactic Center lobe are nonthermal. 

\subsection{Linear Polarization}

A direct and more convincing way to distinguish synchrotron radiation is to 
measure the linear polarization.
However, extremely high Faraday rotation toward the Galactic Center
causes depolarization due to  finite-beam and finite-bandwidth effects.
This  difficulty has been resolved by developing a multi-frequency, 
narrow-band Faraday polarimeter (Inoue et al 1984) as well as by 
high-resolution and high-frequency observations using the VLA 
(Yusef-Zadeh et al 1986). 
Very large rotation measure ($RM > \sim 10^3 $  rad m$^{-2}$) 
and high degree (10 - 50\%) polarization have been observed along the radio 
Arc and in the GCL (Inoue et al 1984;
(Tsuboi et al 1986; Seiradakis et al 1985; Sofue et al 1986; Reich 1988;
 Haynes et al 1992). 

Linar polarization as high as  $p\sim50$ \%  has been detected 
along the Arc at mm wavelengths (Reich et al 1988).
This is nearly equal to the theretical maximum, 
$p_{\rm max}=(\alpha +1)/(\alpha+7/3)\simeq 47$ \%, for the Arc region, where 
the spectral index is  $\alpha\simeq+0.2$.
This implies that the magnetic field is almost perfectly alinged, 
consistent with the VLA observations showing straight filaments
suggestive of highly ordered magnetic field (Yusef-Zadeh et al 1984; 
Morris 1993).  
From linear polarization it is clear that the radio emission from 
the radio Arc is nonthermal despite of its flat radio spectra. 

\section{Radio Continuum Morphology}

\subsection{Themral disk and Star Formation}

The nuclear disk about 50 pc thick and 200 pc in radius comprises numerous 
clumps of HII regions, most of which are active star-forming (SF)
regions, and are detected in the H recombination lines (Mezger and Pauls 1979).
Typical HII regions are named Sgr B, C, D and E.
The total HII mass of $2\times10^6\Msun$ has been estimated, and the 
production rate of Ly continuum photons of $3 \times 10^{52}$ s$^{-1}$ is 
required to maintain this amount of HII gas (Mezger and Pauls 1979).
However, if we take the GC distance of 8.5 kpc and a more accurate
thermal/nonthermal separation, we estimate these to be $\sim 10^6\Msun$
and $1.5\times 10^{52} {\rm s}^{-1}$, respectively.
The SF rate of the central few hundred pc region amounts, therefore,  
to several \% of the total SF rate of the Galaxy.

\subsection{Thermal Filaments}

Complex thermal filaments  connect (bridge) Sgr A with the radio Arc 
(Yusef-Zadeh et al 1984).
Recombination (Pauls et al 1976; Yusef-Zadeh et al 1986) and molecular 
line observations  (G{\"u}sten 1989) indicate their thermal characteristics.
However, large Faraday rotation is detected in the bridge, indicating 
co-existence of magnetic fields along  the thermal filaments
(Sofue et al 1987).
Velocity dispersion of the thermal filamenets
increases drastically near the Arc, indicative of violent
interaction with the Arc (Pauls et al 1976). 
Yusef-Zadeh and Morris (1988) also argue that the Arc 
(straight filaments) and the arched filaments are interacting.
A magneto-ionic jet from Sgr A colliding the ambient poloidal magnetic field
would explain this exotic structure  (Sofue and Fujimoto 1987). 

\subsection{Radio Arc and Vertical Magnetic Fileds}

The radio Arc comprises numerous straight filaments perpendicular 
to the galactic plane, and extends for more than $\sim 100$ pc 
(Downes et al 1978; Yusef-Zadeh  et al 1984; Morris 1993) 
The magnetic field direction is parallel to the filaments and vertical to 
the galactic plane (Tsuboi et al 1986).
Field strength as high as $\sim 1$ mG has been estimated in the Arc and in 
some filaments (Morris 1993).
The life time of cosmic-ray electrons in the Arc is estimated to be  
as short as $\sim 4000$ years (Sofue et al. 1992),
and so the straight filaments  may be transient features, 
temporary illuminated by recently accelerated high-energy electrons.

The higher latitude extension of the Arc, both toward positive and negative 
latitudes, is also  polarized by 10 to 20\%
(Tsuboi et al 1986; Sofue et al 1987). 
The rotation measure reverses across the galactic plane,
indicating a reversal of the line-of-sight component of the magnetic field.
This is consistent with a large-scale poloidal magnetic field 
twisted by the disk rotation (Uchida et al 1985).

\subsection{Galactic Center Lobe and Large-scale Ejection}

The Galactic Center lobe (GCL) is a two-horned vertical structure, probably
a cylinder of about 200 pc in diameter (Sofue and Handa 1984; Sofue 1985: 
Fig. 1). 
The eastern ridge of the lobe is an extension from the radio Arc, and is
strongly polarized. 
The western ridge emerges from Sgr C.
An MHD acceleration model in which the gas is accelerated by a twist of 
poloidal magnetic field by an  accreting gas disk has been proposed (Uchida 
et al 1985; Uchida and Shibata 1986). 
High-velocity molecular gas has been found to be  associated
with the GCL (Sofue 1996: Fig. 1):
Molecular gas in the eastern GCL ridge is receding at 
$\Vlsr \sim +100$ \kms, and the western gas is approaching at 
$\sim -150$ \kms, indicating rotation of the GCL.
This is consistent with the twisted magnetic cylinder model.

\begin{figure}
\vspace{10truecm}
\caption{10 GHz radio map of the Galactic center $2.2\Deg \times 3\Deg$
(left bottom), in comparison with 
a $l-\Vlsr$ plot of the $^{13}$CO emission at $b=8'$ (top). 
showing  high-velocity rotating gas in the GCL.
A VLA map (right)  at 5 GHz shows vertical magnetic
field filaments in the Arc (tick mark interval is $1'$; 
Yusef-Zadeh 1986).
}
\end{figure}

A much larger scale ejection has been found in radio, 
which emanates toward the halo, reaching as high as 
$b\sim 25\Deg$ (Sofue et al 1988).
This feature, which is 4-kpc long and some 200-pc in diameter, may be 
cylindrical in shape and extends roughly perpendicular to the galactic plane.
This structure might be a jet, or it might be magnetic tornado produced by 
the differential rotation between the halo and the nuclear disk.

\subsection{Huge Galactic Bubble by Starburst: North Polar Spur}

The radio North Polar Spur  traces a giant 
loop on the sky of diamter about 120\deg, drawing a huge $\Omega$ over the 
galactic center (Fig. 2).
The $\Omega$-shape can be simulated by  a shock front due to an
explosion (sudden energy input) at the galactic center (Sofue 1994).
In this model, the distance to NPS is several kpc.
The X-ray intensity variation as a function of latitude
indicates that the source is more distant than a few kpc, beyond the
HI gas disk, consistent with the Galactic Center explosion model, but
inconsistent with the local supernova remnant hypothesis.

\begin{figure}
\vspace{8truecm}
\caption{North Polar Spur at 408 MHz after background subtraction 
(Haslam et al1982; Sofue 1994). A shock wave associated with a starburst of 
total energy release of $10^{55}$ ergs is shown at 10, 15 and 20 million years. 
The front at 15 million years can fit the radio shell.
}
\end{figure}

Hence, the NPS is naturally explained, 
if the Galaxy experienced an active phase  15 million years ago
associated with an explosive energy release of some $10^{56}$ erg
(Sofue 1994).
This suggests that  a starburst had occurred in our Galactic Center,
which involved $\sim 10^5$ supernovae during a relatively short period
(e.g., $10^6$ yrs).

\section{Molecular Arms and 120-pc Ring}

Various molecular features have been recoginzed in the central 
$\sim 100-200$ pc region:
such as molecular rings and arms of a few hundred 
pc scale (Scoville et al.1974; Heiligman 1987), 
shell structures and complexes around HII regions (e.g., Hasegawa
et al 1993), and an expanding molecular ring of 200 pc radius
(Scoville 1972; Kaifu et al.1972).
Binney et al. (1991) have noticed a ``parallelogram''
instead of an expanding ring, and interpreted it in terms of  
non-circular kinematics of gas by in an oval potential.
However,  the gas in this parallelogram shares 
only a minor fraction of the total gas mass ($\sim 10$ \%).

\begin{figure}
\vspace{10truecm}
\caption{$^{13}$CO intensity map (Bally et al 1987; Sofue 1995), showing a
highly-tilted ring structure of the molecular gas (bottom), 
in comparison with an \lv\ diagram  at $b=2'$ and $-5'$ (top).  
Arm I and II correspond to the upper and lower parts of the tilted ring.
}
\end{figure}

Fig. 3 shows the total intensity map integrated over the full range of
the velocity (Bally et al 1987; Sofue 1995).
The total molecular mass in the $|l|<1\Deg$ region is estimated to be
$\sim  4.6 \times 10^7 \Msun$ for a new conversion factor (Arimoto et al 1995).
The  molecular mass of the ``disk'' component
is  $\sim 3.9\times10^7 \Msun$, which is 85\% of the
 total in the observed region.
The expanding ring (or the parallelogram)
shares the rest, only $7\times 10^6 \Msun$ (15\%) in the region.

The HI mass within the  central 1 kpc is only of 
several $10^6 \Msun$ (Liszt \& Burton 1980).
Hence,  the central region is dominated
by a molecular disk of $\sim 150$ pc ($\sim 1\Deg$) radius, outside
of which the gas density becomes an order of magnitude smaller.
The total gas mass within 150 pc is only a few percent
of the dynamical mass ( 
$M_{\rm dyn}=R V_{\rm rot}^2/G \sim 8 \times 10^8 \Msun$ for a radius
$R \sim 150$ pc and  rotation velocity $V_{\rm rot} \sim 150$ \kms.
This implies that the self-gravity of gas is not essential in the
galactic center.

Fig. 3 shows \lv\ diagrams in the galactic plane, where 
the foreground components have been subtracted (Sofue 1995).
The major structures of the ``disk component'' near the galactic plane 
are  ``rigid-rotation'' ridges, which we call  ``arms''.
The most prominent arm is found as a long and straight 
ridge, slightly above the galactic plane at $b\sim 2'$, marked
as Arm I in the figure.
Its positive longitude part is connected to the dense molecular
complex Sgr B.
Another prominent arm is seen at negative latitude at $b\sim -6'$,
marked as Arm II.

Arms I and II compose  a bent ring of radius 120 pc with an 
inclination $85\Deg$.
We call this ring the 120-pc Molecular Ring.
It is possible to deconvolve the \lv\ diagram into a 
spatial distribution in the galactic plane by assuming 
approximately circular rotation and using the
velocity-to-space transformation (VST).
Fig. 4 shows a  thus-obtained possible ``face-on'' map of the molecular gas
for  $V_0=150$ \kms.
HII regions are also plotted, showing that HII regions lie  along 
the molecular complexes along the arms in the ring.

\begin{figure}
\vspace{8truecm}
\caption{
 Possible deconvolution of the CO \lv\ diagrams for Galactic Center
 into a face-on view.
}
\end{figure}

\section{Discussion}

We have reviewed various features in the central 150 pc region of the
Galaxy in radio continuum, both in thermal and nonthermal, and in 
the CO line.
The thermal radio emission and molecular gas are distributed within
a thin disk of 150 pc radius and 30 pc thickness.
The nonthermal radio emission is distributed in a wider area, often
extending far from the galactic plane,
comprising vertical structures associated with poloidal magnetic
fields.

We  estimate some energetics among the variuos ISM in the central 150 pc
region:
Molecular gas has turbulent energy density of a few $ 10^{-8}$ erg cm$^{-3}$,
when averaged in the central 150 pc radius disk.
Energy densities due to HII gas and Ly continuum photon are of the order of
 $\sim 10^{-10}$ erg cm$^{-3}$. 
On the other hand, the magnetic field energy is as high as 
$\sim 4\times 10^{-8}$ erg cm$^{-3}$ for $\sim$mG field strength.
The molecular gas disk and magnetic field appear to be in an energy
balance with each other, with the magnetic energy dominating.
The averaged  star formation rate compared to the molecular gas 
amount, or the SF efficiency, is much lower than that observed 
in the outer disk of the Galaxy.
If the Galaxy had experienced a starburst 15 million years 
ago such as that suggested for the cause  of the hugh NPS shell, 
the present Galactic Center may be in a quiet phase, 
probably in a pumping-up phase for the next burst.

\v\v\noindent{\bf References}\v

\r Arimoto, N., Sofue, Y., Tsujimoto, T. 1994, submitted to ApJ.

\r{Bally, J., Stark, A.A., Wilson, R.W., Henkel, C. 1988, ApJ 324, 223.}

\r Binney, J.J., Gerhard, O.E., Stark, A.A., Bally, J., Uchida, K.I., 1991 
MNRAS 252, 210.

\ref{Downes, D.,  Goss, W.M.,  Schwarz, U.J.,  Wouterloot, J.G.A. 1978, 
AA Suppl,  {\bf 35},  1}

\ref{G{\"u}sten, R. 1989,  in {\it The Center of the  Galaxy},  ed. 
M. Morris  (\kluwer),  p.89}

\ref{Haslam, C.G.T.,  Salter, C.J.,  Stoffel, H.,  Wilson, W.E.,  1982,  AA Suppl, 47,  1}

\r Haynes, R. F. ,, Stewart, R. T., Gray, A. D., Reich, W., Reich, P., Mebold, U. 1992, AA 

\r Hasegawa, T., Sato, F., Whiteoak, J. B., Miyawaki, R. 1993, ApJ 419, L77.

\r Heiligman, G. M. 1987 ApJ 314, 747.

\ref{Kaifu, N.,  Kato, T.,  Iguchi, T. 1972,  Nature,  238,  105}

\r Liszt, H. S., Burton, W. B. 1980 ApJ 236, 779.

\ref{Mezger, P.G.,  Pauls, T. 1979,  in {\it The Large-scale characteristics 
of the lGalaxy,  IAU Symp. No.84},  ed. W.B.Burton (D.Reidel,  Drodrecht),  
p.357}

\ref{Morris, M. 1993, The Nuclei of Normal Galaxies, ed. 
R. Genzel \& A.I.Harris, \kluwer), pp..

\ref{Morris, M.,  Yusef-Zadeh, F. 1985,  AJ,  90,  2511}

\ref{Pauls, T.,  Downes, D.,  Mezger, P.G.,  Churchwell, W. 1976,  AA, 46,  407}

\ref{Reich, W.,  Sofue, Y.,  F{\"u}rst, E. 1987,  PASJ, 39,  573}

\r{Scoville, N.Z. 1972, ApJ  175, L127.}

\ref{Seiradakis, J.H.,  Lasenby, A.N.,  Yusef-Zadeh, F.,  Wielebinski, R.,  
Klein, U. 1985,  Nature, 17,  697}

\ref{Shibata, K.,  Uchida, Y. 1987,  PASJ, 39,  559}

\ref{Sofue, Y. 1985,  PASJ,  37,  697}

\r Sofue, Y. 1989, in The Center of the Galaxy (IAU Symp. 136),
ed. M.Morris (D.Reidel Publ. Co., Dordrecht) p. 213.

\r Sofue, Y. 1994, ApJ, 431, L91.

\r Sofue, Y. 1995, PASJ 

\r Sofue, Y. 1996, ApJ. L. in press.

\ref{Sofue, Y.,  Fujimoto, M. 1987,  Ap.J.L, 319,  L73}

\ref{Sofue, Y.,  Handa, T. 1984,  Nature,  310,  568}

\r Sofue, Y., Murata, Y., Reich, W. 1992, PASJ, 44, 367.

\ref{Sofue, Y.,  Reich, W.,  Inoue, M.,  Seiradakis, J.H. 1987,  PASJ, 39, 359}

\ref{Tsuboi, M.,  Inoue, M.,  Handa, T.,  Tabara, H.,  Kato, T.,  Sofue, Y., 
Kaifu, N. 1986,  AJ, 92,  818}

\ref{Uchida, Y.,  Shibata, K. 1986, PASJ, 38,  }

\ref{Uchida, Y.,  Shibata, K.,  Sofue, Y. 1985, Nature, 317, 699}

\ref{Yusef-Zadeh, F.,  Morris, M. 1988,  ApJ, 326,  574}

\ref{Yusef-Zadeh, F.,  Morris, M.,  Chance, D. 1984, Nature 310, 557}

\end{document}